\documentclass{PoS}

\usepackage{multirow}

\newcommand{\Dash}{\boldmath $-$}  

\newcommand{\al}{\alpha}
\newcommand{\be}{\beta}

\newcommand{\de}{\delta}
\newcommand{\De}{\Delta}

\newcommand{\eps}{\epsilon}

\newcommand{\ka}{\kappa}

\newcommand{\beq}{\begin{equation}}
\newcommand{\eeq}{\end{equation}}
\newcommand{\ba}{\begin{array}}
\newcommand{\ea}{\end{array}}
\newcommand{\bea}{\begin{eqnarray}}
\newcommand{\eea}{\end{eqnarray}}
\newcommand{\bi}{\begin{itemize}}  
\newcommand{\ei}{\end{itemize}}
\newcommand{\ben}{\begin{enumerate}} 
\newcommand{\een}{\end{enumerate}}
\newcommand{\bc}{\begin{center}}
\newcommand{\ec}{\end{center}}

\newcommand{\<}{\langle} 
\renewcommand{\>}{\rangle} 
\newcommand{\txt}{\textstyle}

\newcommand{\half} {{\txt \frac{1}{2}}}

\newcommand{\MeV}{{\rm MeV}} 
\newcommand{\fm}{{\rm fm}}

\title{Color superconductivity in ultra-dense quark matter}

\ShortTitle{Color superconductivity in ultra-dense quark matter}

\author{\speaker{Mark Alford}\\
        Physics Department, Washington University, St.~Louis, MO~63130, USA\\
        E-mail: \email{alford@wuphys.wustl.edu},
}

\abstract{
At ultra-high density, matter is expected to form a degenerate
Fermi gas of quarks in which there is
a condensate of Cooper pairs of quarks near the Fermi surface.
This phenomenon is called
color superconductivity. In these proceedings I review 
the underlying physics of color superconductivity and
our current understanding of the possible phases of real-world
quark matter. Then I consider how lattice gauge theorists would
proceed to investigate the phase structure of dense quark matter if 
it were possible to perform the path integral numerically, i.e.~if
the sign problem had been solved.
}

\FullConference{XXIVth International Symposium on Lattice Field Theory\\
                July 23-28, 2006\\
                Tucson, Arizona, USA}

\begin{document}

\section{Introduction}
\label{sec:intro}

The exploration of the phase diagram of matter at ultra-high temperature
or density is an area of great interest and activity, both on the
experimental and theoretical fronts. 
Heavy-ion colliders such as
the SPS at CERN and RHIC at Brookhaven have probed the high-temperature
region, creating and studying the properties of quark matter with very high 
energy density and very low baryon number
density similar to the fluid which filled the universe for 
the first microseconds after the big bang.
Lattice gauge theory calculations have located the critical temperature
and shown that the quark gluon plasma (QGP) is still strongly coupled at
temperatures in heavy ion collisions. There has also been
striking progress in performing lattice calculations in
the ``not too dense'' region, $\mu<T$,
In these proceedings, however, I discuss a different part of
the phase diagram, the low-temperature high-density region where $T\ll\mu$.
Here there are few experimental constraints, and the sign problem
has blocked lattice QCD calculations. However, as I will explain,
we have reasons to expect interesting phase structure.

\subsection{Review of color superconductivity}

QCD is asymptotically free---the interaction becomes weaker as the
momentum grows---so at sufficiently high density and low temperature,
there is a  Fermi surface of almost free quarks. 
The interactions between
quarks near the Fermi surface are certainly attractive in some channels
(quarks bind together to form baryons)
and it was shown by Bardeen, Cooper, and
Schrieffer (BCS) \cite{BCS} that if there is {\em any} channel in which the
interaction is attractive, then there is a state
of lower free energy than a simple Fermi surface. That state arises
from  a complicated coherent 
superposition of pairs of particles (and holes)---``Cooper pairs''.

We can understand the BCS mechanism in an intuitive way as follows.
The Helmholtz free energy is $F= E-\mu N$, where $E$ is
the total energy of the system, $\mu$ is the chemical potential, and
$N$ is the number of fermions. The Fermi surface is defined by a
Fermi energy $E_F=\mu$, at which the free energy is minimized, so
adding or subtracting a single particle costs zero free energy. 
Now switch on a weak attractive interaction.
It costs no free energy to
add a pair of particles (or holes), and the attractive
interaction between them then lowers the free energy of the system.
Many such pairs will therefore
be created in the modes near the Fermi surface, and these pairs,
being bosonic, will form a condensate. The ground state will be a
superposition of states with all numbers of pairs, breaking the
fermion number symmetry. 

Since pairs of quarks cannot be color singlets, the resulting
condensate will break the local color symmetry $SU(3)_{\rm color}$.
We call this ``color superconductivity'' \cite{Reviews}.  Note that
the quark pairs play the same role here as the Higgs particle does in
the standard model: the color-superconducting phase can be thought of
as the Higgs phase of QCD.

\section{The phases of quark matter}
\label{sec:phases}

Quarks, unlike electrons, have color and flavor as well as spin
degrees of freedom, so many different patterns of pairing are possible.
This leads us to expect a rich structure of different color superconducting
phases in quark matter at very high density.

\begin{figure}[t]
\parbox{0.44\hsize}{
\begin{center} \underline{Conjectured form} \end{center}
 \includegraphics[width=\hsize]{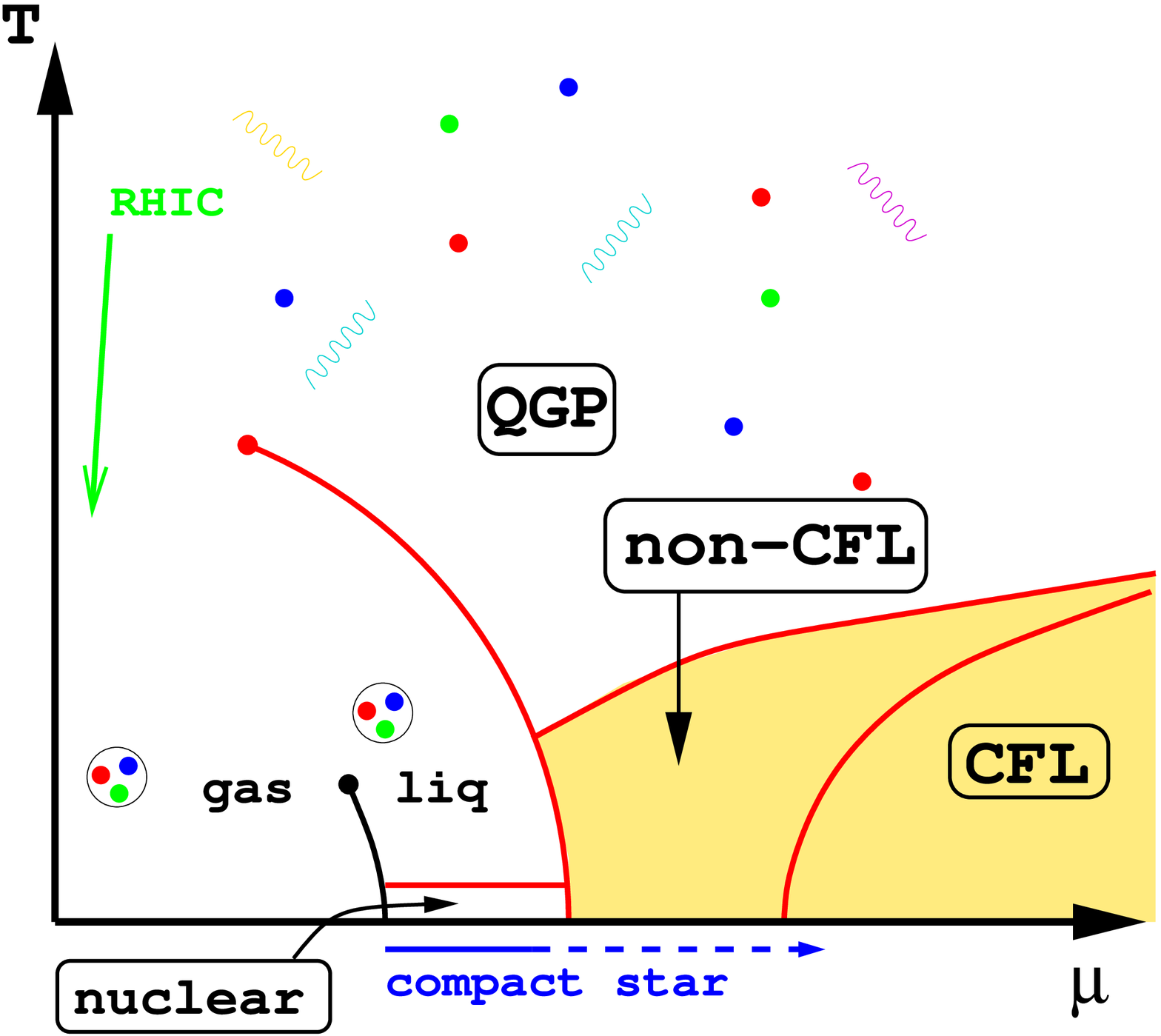}
}
\parbox{0.54\hsize}{
\begin{center} \underline{NJL calculation} \end{center}
 \includegraphics[width=\hsize]{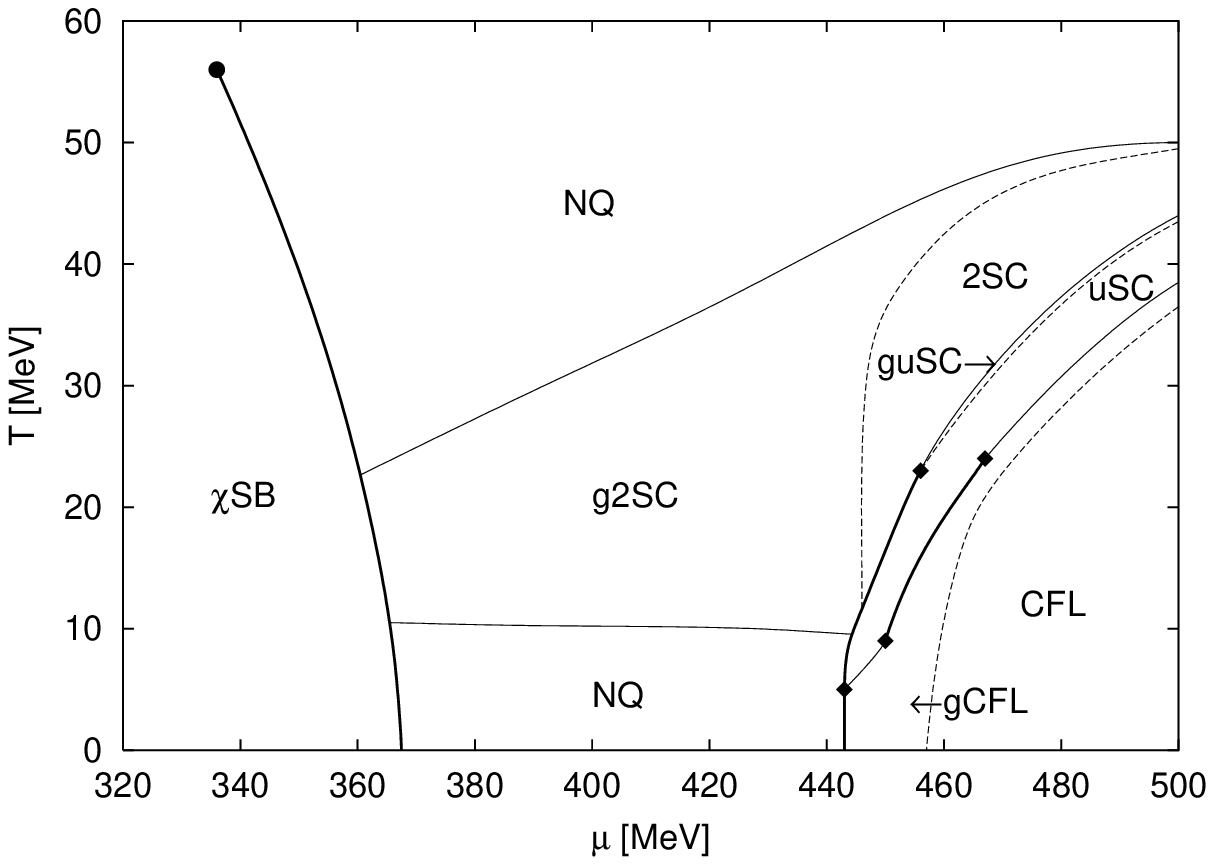}
}
\caption{On the left, the conjectured form of the phase diagram for matter
at ultra-high density and temperature.
On the right, the result of a calculation using an NJL model
\cite{Ruster:2005jc}.
At high density the model has
a rich structure of color-superconducting phases.
Note that the gapless phases (``gCFL'', ``g2SC'' etc) are unstable
(see text).
}
\label{fig:phase}
\end{figure}

In the real world there are two light quark flavors, the up 
($u$) and down ($d$), with 
masses $\lesssim 5~{\rm MeV}$, and a medium-weight flavor, the strange 
($s$) quark, with mass $\sim 90~{\rm MeV}$. (Their effective
``constituent'' masses in dense matter may be much larger.)
The strange quark therefore plays a crucial role in the phases of QCD.
Fig.~\ref{fig:phase} shows a conjectured phase diagram for QCD,
and also a calculated phase diagram obtained using a
Nambu--Jona-Lasinio model of QCD \cite{Ruster:2005jc}.
In both cases, along the horizontal axis the temperature is zero, and
the density rises from the onset of nuclear matter through the
transition to
quark matter. Compact stars are in this region of the phase diagram,
although it is not known whether their cores are dense enough
to reach the quark matter phase.
Along the vertical axis the temperature rises, taking
us through the crossover from a hadronic gas to the quark gluon plasma.
This is the regime explored by high-energy heavy-ion colliders.

At the highest densities we find the CFL phase, in which the strange
quark participates symmetrically with the up and down quarks in Cooper
pairing---this is described in more detail below. It is not yet clear
what happens at intermediate density, and in the next section I will
briefly survey the phases that have been hypothesized to occur there.
The Nambu--Jona-Lasinio model is only a semi-quantitative guide to the
possible behavior of QCD, so its predictions shown in
Fig.~\ref{fig:phase} should be taken as purely illustrative.

\subsection{Highest density: Color-flavor locking (CFL)}
\label{sec:CFL}

It is by now well-established that at sufficiently high densities,
where the up, down and strange quarks can be treated
on an equal footing and the disruptive effects of the
strange quark mass can be neglected, quark matter
is in the color-flavor locked (CFL) phase, in which
quarks of all three colors and all three flavors form
conventional Cooper pairs with zero total momentum,
and all fermionic
excitations are gapped, with the gap parameter 
$\Delta_0\sim 10-100$~MeV \cite{CFL,Reviews}.
This has been confirmed by both NJL \cite{CFL,Schafer:1999pb} and 
gluon-mediated interaction calculations \cite{Schafer:1999fe}.
The CFL pairing pattern is
\begin{equation}
\begin{array}{c}
\langle q^\alpha_i C \gamma_5 q^\beta_j \rangle^{\phantom\dagger}_{1PI}
\propto
  (\kappa+1)\delta^\alpha_i\delta^\beta_j 
+ (\kappa-1) \delta^\alpha_j\delta^\beta_i 
 = \eps^{\al\be N}\eps_{ij N} + \ka(\cdots)  \\[2ex]
 {[SU(3)_{\rm color}]}
 \times \underbrace{SU(3)_L \times SU(3)_R}_{\displaystyle\supset [U(1)_Q]}
 \times U(1)_B 
 \to \underbrace{SU(3)_{C+L+R}}_{\displaystyle\supset [U(1)_{{\tilde Q} }]} 
  \times \mathbb{Z}_2
\end{array}
\end{equation}
Color indices $\alpha,\beta$ and flavor indices $i,j$ run from 1 to 3,
Dirac indices are suppressed,
and $C$ is the Dirac charge-conjugation matrix.
The term multiplied by $\kappa$ corresponds to pairing in the
$({\bf 6}_S,{\bf 6}_S)$, which
although not energetically favored
breaks no additional symmetries and so
$\kappa$ is in general small but not zero 
\cite{CFL,Schafer:1999fe,Shovkovy:1999mr,Pisarski:1999cn}.
The Kronecker deltas connect
color indices with flavor indices, so that the condensate is not
invariant under color rotations, nor under flavor rotations,
but only under simultaneous, equal and opposite, color and flavor
rotations. Since color is only a vector symmetry, this
condensate is only invariant under vector flavor+color rotations, and
breaks chiral symmetry. The features of the CFL pattern of condensation are
\begin{itemize}
\setlength{\itemsep}{-0.7\parsep}
\item[\Dash] The color gauge group is completely broken. All eight gluons
become massive. This ensures that there are no infrared divergences
associated with gluon propagators.
\item[\Dash]
All the quark modes are gapped. The nine quasiquarks 
(three colors times three flavors) fall into an ${\bf 8} \oplus {\bf 1}$
of the unbroken global $SU(3)$, so there are two
gap parameters. The singlet has a larger gap than the octet.
\item[\Dash] 
A rotated electromagnetism (``${\tilde Q} $'')
survives unbroken. Its gauge boson is a combination
of the original photon and one of the gluons.
\item[\Dash] Two global symmetries are broken,
the chiral symmetry and baryon number, so there are two 
gauge-invariant order parameters
that distinguish the CFL phase from the QGP,
and corresponding Goldstone bosons which are long-wavelength
disturbances of the order parameter. 
When the light quark mass is non-zero it explicitly breaks
the chiral symmetry and gives a mass
to the chiral Goldstone octet, but the CFL phase is still
a superfluid, distinguished by its spontaneous breaking of
baryon number.
\item[\Dash]
The symmetries of the
3-flavor CFL phase are the same as those one might expect for 3-flavor
hypernuclear matter \cite{Schafer:1999pb}, so it is possible that there is
no phase transition between them.
\end{itemize}

\section{Real-world intermediate-density quark matter}
\label{sec:real-world}


\begin{figure}
\parbox[t]{0.33\hsize}{
\bc Unpaired\ec
\includegraphics[height=0.8\hsize]{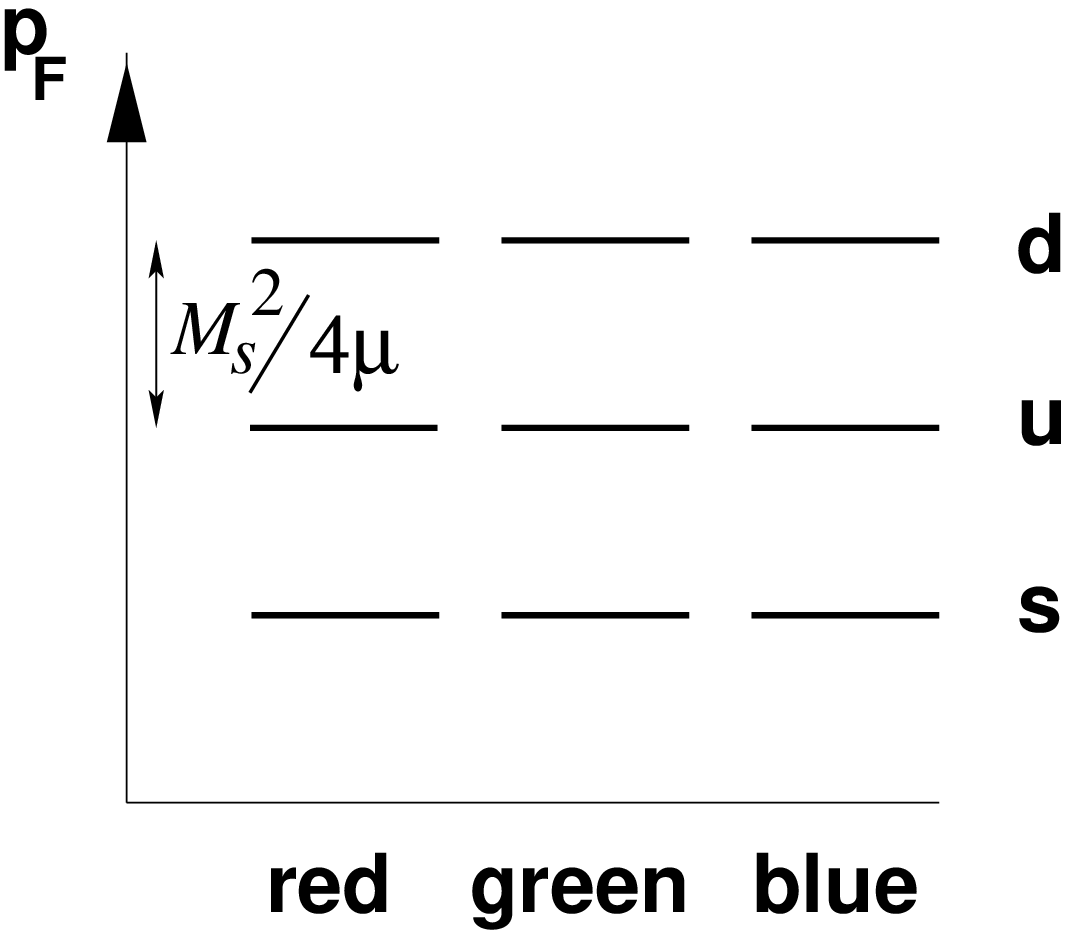}
}\parbox[t]{0.33\hsize}{
\bc 2SC pairing\ec
\includegraphics[height=0.8\hsize]{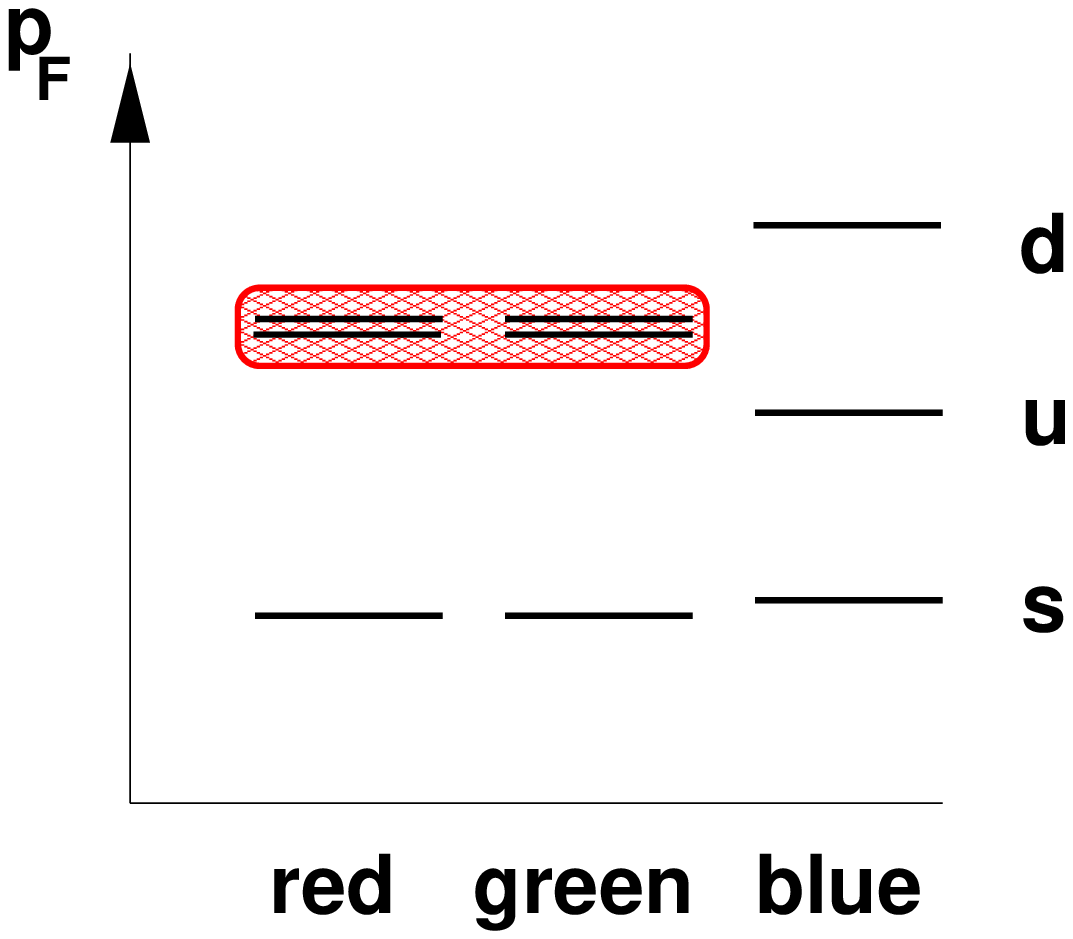}
}\parbox[t]{0.33\hsize}{
\bc CFL pairing\ec
\includegraphics[height=0.8\hsize]{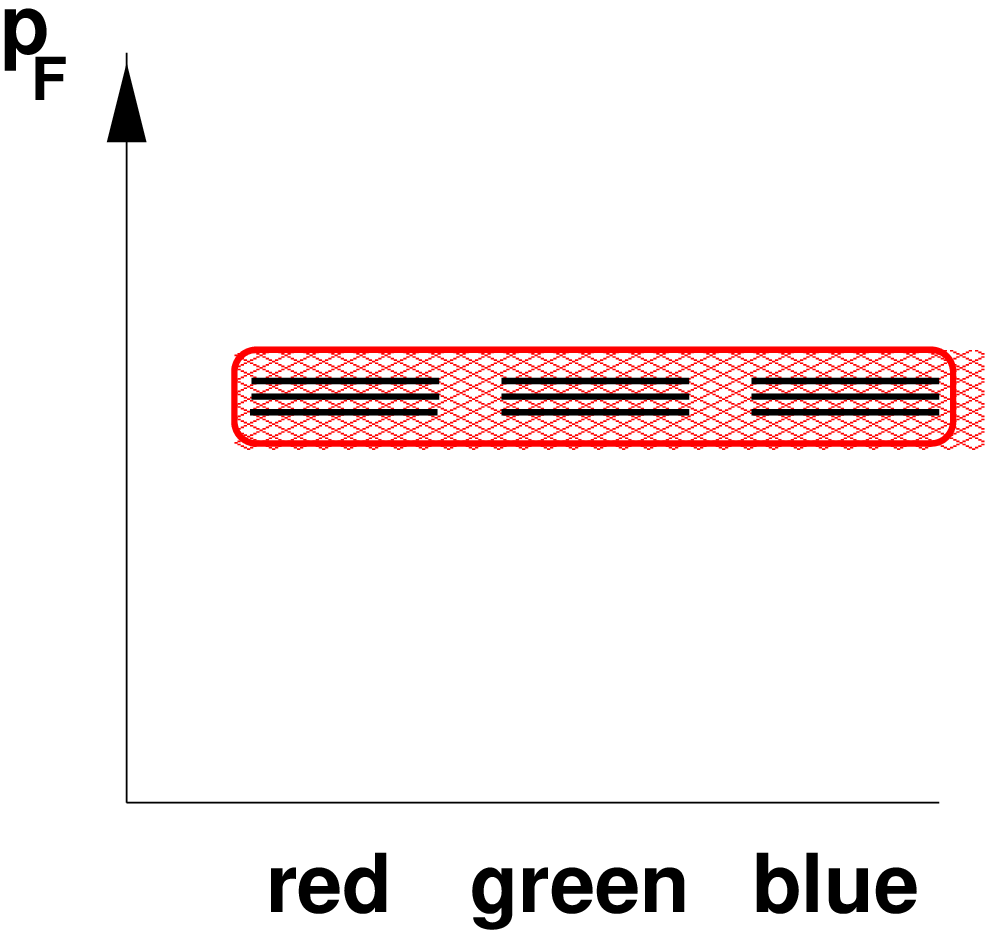}
}
\caption{
Illustration of the splitting apart of Fermi momenta of the
various colors and flavors of quarks. In the unpaired phase,
requirements of neutrality and weak interaction equilibration
cause separation of the Fermi momenta of the various flavors.
In the 2SC phase, up and down quarks of two colors pair,
locking their Fermi momenta together. In the CFL phase,
all colors and flavors pair and have a common Fermi momentum.
}
\label{fig:splitting}
\end{figure}

\subsection{Stresses on the CFL phase}
\label{sec:stresses}

The CFL phase is characterized by pairing between different flavors
and different colors of quarks. This is favored because
the QCD interaction between two
quarks is most attractive in the channel that
is antisymmetric in color (the $\bar{\bf 3}$), and pairing
tends to be stronger in channels that do not break rotational symmetry
\cite{IwaIwa,Schafer:2000tw,Buballa:2002wy,Alford:2002rz,Schmitt:2002sc},
so we expect the pairing to be a spin singlet,
i.e.~antisymmetric in spin.
Fermionic antisymmetry of the Cooper pair wavefunction then
forces the Cooper pair to be antisymmetric in flavor.

Pairing between different colors/flavors can occur easily when
they all have the same chemical potentials and Fermi momenta.
This is the situation at very high density, where the strange quark
mass is negligible.
However, in a real compact star we must take into account the forces that
try to split those Fermi momenta apart, imposing an
energy cost on cross-species pairing. We must
require electromagnetic and color
neutrality \cite{Iida:2000ha,Alford:2002kj} (possibly
via mixing of oppositely-charged phases),
allow for equilibration under the weak interaction, and include a
realistic mass for the strange quark.  
These factors cause the different
colors and flavors to have different chemical potentials, and
this imposes a stress on cross-species pairing
such as occurs in the CFL pairing pattern.
This is illustrated in Fig.~\ref{fig:splitting}, which shows the Fermi
momenta of the different species of quarks. In the unpaired
phase, the strange quarks have a lower Fermi momentum because
they are heavier, and to maintain electrical neutrality the number
of down quarks is correspondingly increased (electrons are also
present, but play a small role in maintaining neutrality).
To lowest order in the strange quark mass, the separation between
the Fermi momenta is $M_s^2/(4\mu)$, so the splitting is more pronounced
at lower density.
If the attraction between quarks is sufficiently strong,
color superconductivity can overcome this splitting of the
Fermi momenta. In the 2SC phase \cite{Alford:1997zt,Rapp:1997zu},
the up and down quarks of two of the
colors undergo Cooper pairing, which locks their Fermi momenta together.
The pairing will only occur if the energy released by the formation
of the condensate is greater than the energy cost of moving the
quark Fermi surfaces away from their ``natural'' positions in the
unpaired phase. To lowest order in $M_s$, 2SC pairing is favored 
relative to the unpaired phase when
the smearing of the Fermi surface due to Cooper pairing is greater than
the splitting, i.e. when $\De_{\rm 2SC}> M_s^2/(4\mu)$
\cite{Alford:2002kj} (this estimate assumes that contributions to the
free energy from the chiral condensate are the same in both phases).
In the CFL phase, the pairing is extended to all
colors and flavors, which are then locked together
with a common Fermi momentum, and the criterion for pairing to
occur turns out to be the same, $\De_{\rm CFL}> M_s^2/(4\mu)$.

At ultra-high density the splitting between the Fermi momenta becomes
negligible, and the CFL phase is favored. However, as the density drops
to values that might realistically occur in the core of a neutron star,
the value of $M_s^2/(4\mu)$ rises to tens of MeV, which is
of the same order as the expected pairing gap $\De$ in the
2SC and CFL phases.
Thus as the density decreases we expect the CFL pairing pattern
to be distorted, and then to be replaced by some other
pattern. NJL model calculations
\cite{Fukushima:2004zq,Abuki:2004zk,Blaschke:2005uj,Ruster:2005jc}
find that if the attractive interaction is very strong
(so that $\De_{CFL}\sim 100~\MeV$ where
$\De_{CFL}$ is what the CFL gap would be at $\mu\sim 500~\MeV$
if $M_s$ were zero)
then the CFL phase survives all the way down to the transition
to nuclear matter. If it is a little less strong
then there may be an interval of 2SC phase
but in general the 2SC phase does not offer a better
compromise between pairing and Fermi surface splitting
\cite{Alford:2002kj,Steiner:2002gx}.
A comprehensive survey of possible BCS pairing patterns shows
that all of them suffer from the stress of Fermi surface splitting
\cite{Rajagopal:2005dg}, so in the intermediate-density region
more exotic phases are expected.
In the next few subsections we give a quick overview of the
expected phases of real-world quark matter at intermediate density. 
We restrict our
discussion to zero temperature because the critical temperatures
for most of the phases that we discuss are expected to be
of order $10~\MeV$ or higher, and the core temperature
of a neutron star is believed to drop below this value
within minutes (if not seconds) of its creation in a supernova.

\subsection{Kaon condensation: the CFL-$K^0$ phase}
Bedaque and Sch\"afer 
\cite{BedaqueSchaefer} showed that when the stress is not too large 
(high density), it may  simply
modify the CFL pairing pattern by inducing a flavor rotation of
the condensate which can be interpreted as a condensate
of ``$K^0$'' mesons, i.e.~the neutral anti-strange Goldstone bosons
associated with the chiral symmetry breaking. 
This is the ``CFL-$K^0$'' phase, which breaks isospin.
The $K^0$ condensate can easily be suppressed by
instanton effects \cite{Schafer:2002ty}, but if these are
ignored then the kaon condensation occurs for
$M_s \gtrsim m^{1/3}\De^{2/3}$ for light ($u$ and $d$) quarks
of mass $m$.
This was demonstrated
using an effective theory of the Goldstone bosons, but
with some effort can also be seen in an NJL calculation
\cite{Buballa:2004sx,Forbes:2004ww}.

\subsection{The unstable gapless phases}
\label{sec:gCFL}
\begin{figure}[t]
 \bc
 \includegraphics[width=0.7\hsize]{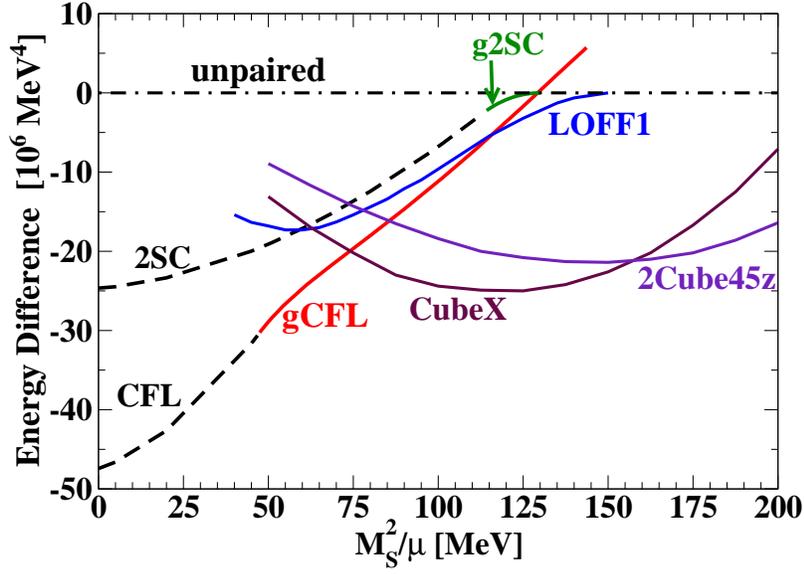}
 \ec
\caption{
Free energy of various phases of dense QCD.
The CFL pairing strength is $\De_{CFL}=25~\MeV$. The curves for
the CFL, 2SC, gCFL, g2SC, and LOFF1 phases are obtained from an
NJL model.
Note that the gCFL phase takes over
from CFL at $M_s^2/\mu \approx 2 \De_{CFL}$, and remains favored
beyond the value $M_s^2/\mu \approx 4 \De_{CFL}$ at which the
CFL phase would become unfavored. The ``LOFF1'' curve is the
single-plane-wave LOFF ansatz of \cite{Casalbuoni:2005zp}.
The ``CubeX'' and ``2Cube45z'' lines are estimates for more complicated
LOFF crystal structures, and follow from the Ginzburg-Landau
calculation of Ref.~\cite{Rajagopal:2006ig}.
}
\label{fig:energy}
\end{figure}

The NJL analysis shown in Fig.~\ref{fig:phase} predicts 
that at densities too low for CFL pairing there will be
gapless phases (``gCFL'',''g2SC'', etc). This can be
understood by a rough quantitative analysis that involves
expanding in powers of
$M_s/\mu$ and $\De/\mu$, and
ignoring the fact that the effective strange
quark mass may be different in different phases \cite{Alford:2002kj}.
Such an analysis shows that as we come down in density 
we find a transition at $\mu \approx \half M_s^2/\De_{CFL}$
from CFL to another phase, the gapless CFL phase (gCFL) \cite{Alford:2003fq}.
The underlying physics here is that when $\mu<\half M_s^2\De_{CFL}$
it becomes energetically favorable to convert a
$gs$ quark near the common Fermi momentum into a $bd$ quark,
breaking the Cooper pairing over a range of momenta in that channel.
The free energies of the competing phases in an NJL model
are shown in Fig.~\ref{fig:energy}:
The gCFL phase takes over
from CFL at $M_s^2/\mu \approx 2 \De_{CFL}$, and remains favored
beyond the value $M_s^2/\mu \approx 4 \De_{CFL}$ at which the
CFL phase would become unfavored.

However, it turns out that the gapless phases are unstable.  The
instability of the gCFL phase was established in
Refs.~\cite{Casalbuoni:2004tb,Fukushima:2005cm} after an analogous
instability in the gapless 2SC phase had been discovered
\cite{Huang:2004am,Giannakis:2004pf}. The instability manifests itself
in an imaginary Meissner mass $M_M$ for some of the gluons.  $M_M^2$
is the low-momentum current-current two-point function, and
$M_M^2/(g^2\De^2)$ (where the strong interaction coupling is $g$)
is the coefficient of the gradient term in the
effective theory of small fluctuations around the ground-state
condensate.  The fact that we find a negative value when the
quasiparticles are gapless indicates an instability towards
spontaneous breaking of translational invariance
\cite{Reddy:2004my,Huang:2005pv,Iida:2006df,Hashimoto:2006mn,Fukushima:2006su}.
Calculations in a simple two-species model \cite{Alford:2005qw} show
that gapless charged fermionic modes generically lead to imaginary
$M_M$.

The instability of the gapless phases indicates that
there must be 
other phases of even lower free energy, that occur in their place
in the phase diagram. The nature of those phases remains uncertain at 
present: some candidates are discussed below.

\subsection{Crystalline pairing}

The pairing patterns discussed so far have been translationally
invariant. But in the region of parameter space where cross-species
pairing is just barely excluded by stresses that pull apart the Fermi
surfaces, one expects a position-dependent pairing known as the
``LOFF'' phase
\cite{LOFF,Alford:2000ze,Bowers:2002xr,Casalbuoni:2003wh}.  This
arises because one way to achieve pairing between different flavors
while accommodating the tendency for the Fermi momenta to separate is
to only pair over part of the Fermi surface, and form pairs with 
non-zero momentum.  The LOFF phase therefore
competes with the gCFL phase, and may resolve that phase's stability
problems.

Recent calculations for 3-flavor quark matter (within a
Ginzburg-Landau approximation) show that even a very
simple single-plane-wave LOFF ansatz yields a state that has lower
free energy than gCFL in the region where the gCFL$\to$unpaired
transition occurs \cite{Casalbuoni:2005zp,Casalbuoni:2006zs} (see
Fig.~\ref{fig:energy}) and the Meissner instability is no longer
present \cite{Ciminale:2006sm}.

The gap parameter and free energy for three-flavor quark matter
have also recently been evaluated
within a Ginzburg-Landau approximation for many candidate crystal
structures \cite{Rajagopal:2006ig}.  
Fig.~\ref{fig:energy} shows the free energies of the two
most favorable crystal structures, CubeX and 2Cube45z.
The robustness of these phases results in their being favored over wide
ranges of density.  However, it also implies that the Ginzburg-Landau
approximation is not quantitatively reliable, so the CubeX and 2Cube45z
lines in Fig.~~\ref{fig:energy} should only be taken as an indication that
the LOFF state might be preferred to gCFL
over a much wider range of the stress parameter $M_s^2/(2\mu\De)$
than one would infer from the single plane wave calculation.

\subsection{Meson supercurrent (``curCFL'')}

Kaon condensation alone does not remove the gapless modes
that occur in the CFL phase when $M_s$ becomes
large enough. The CFL-$K^0$ phase also develops gapless modes
and a Meissner instability, though at a slightly larger value of $M_s$
\cite{Kryjevski:2004jw,Kryjevski:2004kt}. One way that the
CFL-$K^0$ phase can respond is by
developing a current in the pseudo-Goldstone bosons (kaons),
i.e. a spatial modulation of the $K^0$ condensate
\cite{Schafer:2005ym,Kryjevski:2005qq}. There is no net transfer
of any charge because there
is a reverse flow in the gapless fermions.
The meson current lowers the free energy, but is essentially just another
instability: as yet there is no analysis that finds a new
meson-supercurrent ground state, whose free energy could be compared
with that of other states such as LOFF phases.
Calculation of the meson supercurrent in the CFL phase (with no
uniform background $K^0$ condensate) shows that
it is induced when gapless quark modes appear, and that it
resolves the Meissner instability, but in that case it is
equivalent to a plane-wave LOFF state \cite{Gerhold:2006dt}.

\subsection{Gluon condensation}

Analysis of the magnetic instability in the two-flavor gapless
color-superconducting phase (g2SC) using a Ginzburg-Landau approach
has found that the instability can be cured by the appearance of a
chromoelectric condensate \cite{Gorbar:2005rx,Gorbar:2006up}. The 2SC
condensate breaks the color group down to the $SU(2)_{rg}$ red-green
subgroup, and five of the gluons become massive vector bosons via the
Higgs mechanism. The new condensate involves some of these massive
vector bosons, and because they transform non-trivially under
$SU(2)_{rg}$ it now breaks that gauge symmetry. Because they are
electrically charged vector particles, rotational symmetry is also broken, 
and the phase is an electrical superconductor.
There are some connections between the gluon condensate and the LOFF
phase: the single-plane-wave LOFF state is gauge-equivalent to
a homogeneous vector boson condensate. However in general the gluon
condensate has non-zero field strength, and is not simply a gauge
transformation of an inhomogeneous diquark condensate
\cite{Hashimoto:2006mn}. In the two-flavor case, gluon condensation
appears to be favored over single-plane-wave LOFF \cite{Kiriyama:2006ui},
but it has not been compared with a LOFF crystal,
and as yet the gluon condensate has not been
studied in the three-flavor case.

\subsection{Secondary pairing}

Since the Meissner instability is generically associated with the presence of
gapless fermionic modes, and the BCS mechanism implies that any
gapless fermionic mode is unstable to Cooper pairing in the most
attractive channel, one might expect that the instability will simply
be resolved by ``secondary pairing'' of the gapless quasiparticles
which would then acquire their own gap $\De_s$
\cite{Hong,Huang:2003xd}.  Furthermore, the quadratically gapless mode
in gCFL has a greatly increased density of states at low energy
(diverging as $E^{-1/2}$), so its secondary pairing is much stronger
than would be predicted by BCS theory: $\Delta_s \propto G_s^2$ for
coupling strength $G_s$, as compared with the standard BCS result
$\Delta \propto \exp(-{\rm const}/G)$ \cite{Hong}.
This result is confirmed by an NJL study in a
two-species model \cite{Alford:2005kj}, but the the secondary gap
was found to be still much smaller than
the primary gap, so it does not generically resolve the magnetic
instability (in the temperature range $\De_s \ll T \ll \De_p$, for example).

\subsection{Single-flavor pairing}

At low enough density, $M_s$ puts such a significant stress on the
pairing pattern that no pairing between different flavors
is possible \cite{Rajagopal:2005dg}.  The resultant phase is
often called ``unpaired'' quark matter, but there remains the
possibility of Cooper pairing where each flavor pairs with itself.
(This regime will only arise if $\Delta_0$ is so small that very large
values of $M_s^2/(\mu\Delta_0)$ can arise without $\mu$ being taken so
small that nuclear matter becomes favored.)  Single-flavor pairing may
also arise in the 2SC phase, where the strange quarks are not involved
in two-flavor pairing.
Single-flavor pairing phases have much lower critical temperatures 
than multi-flavor phases like the CFL or 2SC phases, 
perhaps as large as a few MeV, more typically
in the eV to many keV range
\cite{Schafer:2000tw,Buballa:2002wy,Alford:2002rz,Schmitt:2002sc,Schmitt:2004et},
so they are expected to play a role late in the life of a neutron star.

\noindent
$\bullet$ {\em Single flavor pairing in ``unpaired'' quark matter}.
In most NJL studies, matter with no cross-species pairing at all is
described as
as ``unpaired'' quark matter. However, it is well known that
there are attractive channels for a single flavor pairing, although they
are much weaker than the 2SC and CFL 
channels
\cite{Schafer:2000tw,Buballa:2002wy,Alford:2002rz,Schmitt:2002sc,Schmitt:2004et}.
Calculations using NJL models and single-gluon exchange agree
that the favored phase in this case is the color-spin-locked (CSL) 
phase \cite{Schafer:2000tw}
in which there is pairing of
all three colors of each flavor, with each pair of colors
correlated with a particular direction for the spin.
This phase does not break rotational symmetry.



\noindent
$\bullet$ {\em Single flavor pairing in 2SC quark matter}.
If there is a regime in which the 2SC phase survives,
this leaves the blue quarks unpaired.
In that case one might expect a ``2SC+CSL'' pattern, 
which would again be rotationally symmetric,
in which the strange quarks of all three colors
self-pair in the CSL pattern.
However, the 2SC pattern breaks the color symmetry, and
in order to maintain color neutrality, a color chemical potential
is generated, which also affects the unpaired strange quarks,
splitting the Fermi momentum of the blue strange quarks
away from that of the red and green strange quarks
(see middle panel of Fig.\ref{fig:splitting}).
This is a small effect, but so is the CSL pairing gap, 
and NJL model calculations indicate
that the color chemical potential typically destroys CSL 
pairing of the strange quarks \cite{Alford:2005yy}.
The system falls back on the next best alternative, which
is spin-1 pairing of the red and green strange quarks.

\subsection{Mixed Phases}
Another way for a system to deal with a stress on its
pairing pattern is phase separation. In the context of quark
matter this corresponds to relaxing the requirement of local charge
neutrality, and requiring neutrality only over long distances,
so we allow a mixture of a positively charged and a negatively charged
phase, with a common pressure and a common value of the electron
chemical potential $\mu_e$ that is not equal to the neutrality value
for either phase. Such a mixture of nuclear and CFL quark matter
was studied in Ref.~\cite{Alford:2001zr}. In quark matter
it has been found that as long as we require local color neutrality 
such mixed phases are not the favored response to the stress imposed
by the strange quark mass \cite{Alford:2003fq,Alford:2004nf}. Phases involving
color charge separation have been studied \cite{Neumann:2002jm} but it seems
likely that the energy cost of the color-electric fields will disfavor them.

\section{Quark matter on the lattice}

For neutron star phenomenology, including color superconductivity
in quark matter, the relevant part of the phase diagram is
the high-density low-temperature region. Although there has
been great progress in mitigating the effects of the
sign problem in the complementary region of low chemical
potential and high temperature 
\cite{Philipsen:2005mj,Fodor:2004nz,Allton:2005gk}, 
the sign problem remains a
severe obstacle to lattice calculations at $\mu\gg T$.
It should be noted, however, that the sign problem is more
a technical problem than a fundamental one. 
There is no ``no-go'' theorem stating
that we can never expect to perform lattice QCD calculations at
$\mu\gg T$. In certain theories the sign problem has been completely
solved, for example the 3-state Potts model \cite{Alford:2001ug}
and the $O(3)$ model \cite{Chandrasekharan:2001ya},
and work is in progress to extend these methods
to QCD~\cite{Wiese:2005yi}. It is therefore useful
to think about how we would study color superconductivity
using lattice QCD if we could perform lattice QCD calculations
in the high-density low-temperature region of the phase diagram.

In lattice calculations we have the freedom to vary parameters that
are fixed in the real world, such as the number of quark flavors
and their masses. Also, issues of electrical charge neutrality
and equilibration under the weak interactions do not arise, so
lattice QCD would be able to turn off the stresses that were
discussed in section \ref{sec:stresses}. One significant constraint, however,
is the size of the lattice. The size in the Euclidean time direction
corresponds to the temperature, and current calculations are
limited to sizes less than 5~fm, corresponding to $T\geqslant 40~\MeV$.
A $20~\fm$ lattice, which is very large by current standards,
corresponds to $T=10~\MeV$. Of course, a breakthrough that
allows us to work at $\mu\gg T$ may also allow us to work in
very large volumes, but the more conservative assumption is
that it will remain difficult
to study phases with critical temperatures lower than about 10 MeV.
Superfluidity in nuclear matter, with a critical temperature around
1 MeV, will be therefore be completely inaccessible, and
we will have to search 
for phases with higher critical temperatures. Fortunately,
many color superconducting phases are expected to have
appropriately high critical temperatures.

\subsection{Quark Matter with $N_f$ massless flavors}

\begin{table}
\newlength{\rcol}
\setlength{\rcol}{0.35\hsize}
\begin{tabular}{ccll}
$N_f$ & phase & global symmetry group & description \\
\hline
2 & unbroken (QGP):& $\phantom{\to\;} SU(2)_L\otimes SU(2)_R\otimes U(1)_B$  \\
  & vacuum: & $\to SU(2)_V \otimes U(1)_B$ 
     & chiral symmetry broken ($\chi$SB) \\
  & hadronic: & $\to SU(2)_V$ & $\chi$SB and superfluid \\
  & 2SC: & $\to SU(2)_L\otimes SU(2)_R   \otimes U(1)_{\tilde B}$
     & same as QGP \\[1ex]
3 &unbroken (QGP):&  $\phantom{\to\;} SU(3)_L\otimes SU(3)_R\otimes U(1)_B$  \\
  &vacuum: &  $\to SU(3)_V \otimes U(1)_B$ & $\chi$SB \\
  & hadronic: & $\to SU(3)_V$ & $\chi$SB and superfluid \\
  &CFL: &  $\to SU(3)_{L+R+c} \otimes Z_2$ 
     & same as hadronic \\[1ex]
4 &unbroken (QGP):&  $\phantom{\to\;} SU(4)_L\otimes SU(4)_R\otimes U(1)_B$ \\
  &vacuum: &  $\to SU(4)_V \otimes U(1)_B$ & $\chi$SB  \\
  &hadronic: &  $\to SU(4)_V$ & $\chi$SB and superfluid\\
  &P$\chi$SB: &  $\to SU(2)_{V}\otimes SU(2)_V \otimes SU(2)_A$ &
    unique  \cite{Schafer:1999fe} \\
\hline
\end{tabular}
\caption{
Symmetry breaking patterns for various phases of QCD with
2,3, and 4 quark flavors. For each $N_f$ the last entry is the expected
form of color superconductivity at the highest densities.
}
\label{tab:breaking}
\end{table}

In table \ref{tab:breaking} we give the expected global symmetries
of various phases of QCD with $N_f=2,3,4$.
We do not include $N_f=1$
because single-flavor color superconducting phases are predicted
to have critical temperatures of order 1 MeV or less
\cite{Schafer:2000tw,Buballa:2002wy,Alford:2002rz,Schmitt:2002sc},
so they are
not likely to be seen on lattices of a reasonable size.
It is noticeable from table  \ref{tab:breaking} that color superconducting
phases are not easy to identify. In two-flavor quark matter, the
2SC color superconductor leaves all the global symmetries unbroken,
so there is no order parameter that distinguishes it from unpaired quark
matter or quark gluon plasma \cite{Alford:1997zt}.
In three-flavor quark matter, the CFL color superconductor breaks
the global symmetries in exactly the same way as hadronic matter,
including complete breaking of the chiral symmetry and superfluidity
(since all quarks are massless the baryons are all degenerate,
and so the baryon octet can self-pair in a pattern that preserves
the flavor symmetry \cite{Alford:1999pa,Schafer:1999pb}).
In the four-flavor theory, however, Sch\"afer \cite{Schafer:1999fe}
finds an interesting ``partially chirally broken'' (P$\chi$SB)
phase, which has different symmetries from any of the other expected phases.
Since the staggered fermion formalism
naturally yields four continuum flavors, this might be a good place to 
begin the
search for color superconductivity on the lattice.

\subsection{Probing phases and symmetry breaking on the lattice}

There are various tools for to distinguishing different phases
of high-density QCD on the lattice.

\noindent
$\bullet$ {\em Measuring local order parameters}. Technically,
spontaneous symmetry breaking
occurs only in infinite volume systems, where the chance of
making transitions between the different possible vacua is zero.
More practically, we expect to see spontaneous symmetry breaking
when the limit of large
volume is taken first, {\em before} the limit of taking external
currents to zero in the functional integral:
\beq
\< \phi\> =  \lim_{J\to 0}\,\lim_{V\to \infty}\,
  \frac{\de}{\de J} \int\!D\phi \exp(-S[\phi] + J\phi) \ .
\eeq
This delicate procedure has been implemented in simpler theories
such as the Gross-Neveu model \cite{Hands:2002mr}.
The order parameter for superfluidity (breaking of $U(1)_B$)
will be a color and flavor singlet dibaryon. The order parameter
for chiral symmetry breaking could be the conventional 
color and flavor singlet chiral condensate, but it is expected that
this is suppressed relative to a four-fermion operator 
$\bar\psi\bar\psi \psi\psi$ with the
same quantum numbers \cite{Schafer:1999fe}.

\noindent
$\bullet$ {\em Measuring gaps in the fermion spectrum}.
One of the characteristic consequences of Cooper pairing is
the generation of gaps in the fermion spectrum. In color superconducting
phases, therefore, we expect to find that the fermionic excitations
are classified by representations of the unbroken symmetry group,
and that some of them are gapped. In QCD the fundamental
fermions are quarks, but gauge invariance dictates that quark 
``quasiparticles'' are still created by baryon creation operators,
so the procedure for finding the gaps is the same as that for measuring
baryon masses in zero-density QCD.
Again, this pairing signature has been seen in the Gross-Neveu model
\cite{Walters:2003it}, and in the same paper evidence was also found of
particle-hole mixing in the fermion spectrum, which is another
characteristic of Cooper pairing.

\noindent
$\bullet$ {\em Measuring low masses of Goldstone bosons}.
The breaking of a continuous global symmetry, as well as giving a non-zero 
value to some order parameter, creates massless bosonic modes in the
spectrum of the theory, known as Goldstone bosons. For chiral symmetry
breaking these are the pions. The procedure for measuring the masses is
the same as that for measuring meson masses in zero-density QCD.


\section{Conclusion}

As I have described, the project of delineating a plausible phase
diagram for real-world high-density quark matter is still not complete. I have
discussed some ideas for the ``non-CFL'' region of Fig.~\ref{fig:phase}, 
but there are others
such as deformation of the Fermi surfaces
(discussed so far only in non-beta-equilibrated nuclear matter
\cite{Sedrakian:2003tr}) or
a Bose-Einstein condensate (BEC) of spatially-bound diquarks
\cite{Sedrakian:2005db}.
It is very interesting to note that the situation we find
in quark matter, a system with
pairing that must respond to a stress that separates the chemical potentials
of the pairing species, is a very generic one, arising also in
condensed matter systems and cold atom systems \cite{ketterle,hulet}.
Recent work on BCS/BEC crossover in asymmetric dilute Fermi gases 
\cite{Son:2005qx,Gubankova:2006gj,Rupak:2006et,Bulgac:2006gh,Mannarelli:2006hr} 
shows that between the BCS-paired region and the
unpaired region in the phase diagram 
one should expect a translationally-broken
region. In QCD this could correspond to a $p$-wave meson condensate,
a gluon condensate, or a LOFF state. What is
particularly exciting is that the technology of cold atom traps
has advanced to the point where fermion superfluidity can
now be seen in conditions where many of the important parameters
can be manipulated, and it may soon be possible to investigate
the response of the pairing to external stress under
controlled experimental conditions.

My discussion of lattice approaches to color superconductivity
was hypothetical, awaiting a breakthrough that would allow us to
evaluate the functional integral at high density
and low temperature. Current efforts in this direction include
the development of a D-theory formulation that would allow the
application of cluster algorithms \cite{Wiese:2005yi}, and 
also approaches using strong coupling and Hamiltonian methods.
In particular, the effective strong-coupling Hamiltonian 
in the large $N_c$ limit has been written
as an antiferromagnet with next-to-nearest neighbor couplings,
and indications have been found of chiral condensation,
but not as yet of quark Cooper pair condensation
\cite{Bringoltz:2002qc,Bringoltz:2003jf,Luo:2004se,Nishida:2003fb}.

At the moment, then, the study of dense quark matter has yielded a
diverse landscape of possible phases and phenomenologies across which
theorists roam quite freely. Of course there is much interesting work
to be done in exploring this territory.
But it is also to be hoped that in the future,
with increasingly precise observations of neutron star behavior
and perhaps even a leap forward in our ability to perform QCD calculations at
the relevant densities, we will start to close in on the
real geography of the phase diagram of QCD.



\end{document}